\documentclass[conference]{IEEEtran}
\usepackage{cite}
\usepackage{amsmath,amssymb,amsfonts}
\DeclareMathOperator*{\argmax}{arg\,max}
\usepackage{algorithmic}
\usepackage{graphicx}
\usepackage{textcomp}
\usepackage{xcolor}
\usepackage{multirow}
\usepackage{diagbox}
\usepackage{bm}
\usepackage{graphicx}
\usepackage{epstopdf}
\usepackage{tikz}

\newcommand{\C}[1]{}
\newcommand{\dbu}[1]{\ensuremath{#1\,\text{dB}}}

\newcommand{\tm}{\ensuremath{(k-1)}}
\newcommand{\noiflr}{\ensuremath{\sigma_{N}^2}}
\newcommand{\noiflre}{\ensuremath{\widehat{\noiflr}}}
\newcommand{\noiper}{\ensuremath{\mathbb{E}\big(|N|^2\big\rvert X \big)}}

\def\BibTeX{{\rm B\kern-.05em{\sc i\kern-.025em b}\kern-.08em
    T\kern-.1667em\lower.7ex\hbox{E}\kern-.125emX}}

\newcommand\copyrighttext{%
  \footnotesize \textcopyright 2021 IEEE. Personal use of this material is permitted. Permission from IEEE must be obtained for all other uses, in any current or future media, including reprinting/republishing this material for advertising or promotional purposes, creating new collective works, for resale or redistribution to servers or lists, or reuse of any copyrighted component of this work in other works.}
\newcommand\copyrightnotice{%
\begin{tikzpicture}[remember picture,overlay]
\node[anchor=south,yshift=10pt] at (current page.south) {\fbox{\parbox{\dimexpr\textwidth-\fboxsep-\fboxrule\relax}{\copyrighttext}}};
\end{tikzpicture}%
}

\begin{document}

\title{Robust Acoustic Scene Classification in the Presence of Active Foreground Speech}
\author{\IEEEauthorblockN{Siyuan Song, Brecht Desplanques, Celest De Moor, Kris Demuynck, Nilesh Madhu}
\IEEEauthorblockA{IDLab, Department of Electronics and Information Systems, Ghent University - imec, Belgium\\
Email: \{Siyuan.Song, Brecht.Desplanques, Celest.Demoor, Kris.Demuynck, Nilesh.Madhu\}@ugent.be}}
\maketitle
\copyrightnotice

\begin{abstract}
We present an iVector based Acoustic Scene Classification (ASC) system suited for real life settings where active foreground speech can be present. In the proposed system, each recording is represented by a fixed-length iVector that models the recording's important properties.
A regularized Gaussian backend classifier with class-specific covariance models is used to extract the relevant acoustic scene information from these iVectors.
To alleviate the large performance degradation when a foreground speaker dominates the captured signal, we investigate the use of the iVector framework on Mel-Frequency Cepstral Coefficients (MFCCs) that are derived from an estimate of the noise power spectral density.
This noise-floor can be extracted in a statistical manner for single channel recordings. We show that the use of noise-floor features is complementary to multi-condition training in which foreground speech is added to training signal to reduce the mismatch between training and testing conditions.
Experimental results on the DCASE 2016 Task~1 dataset show that the noise-floor based features and multi-condition training realize significant classification accuracy gains of up to more than 25 percentage points (absolute) in the most adverse conditions. These promising results can further facilitate the integration of ASC in resource-constrained devices such as hearables.
\end{abstract}


\section{Introduction}

Enhancement of audio and speech is typically a first stage in many audio applications, with a view to improving the user experience.
For optimal results, the parameters of the enhancement approaches should be adaptive to the environment. To enable such advanced techniques, the nature of the background environment should be determined first. The classification of the background audio into acoustic {\em scenes} is termed as Acoustic Scene Classification (ASC) \cite{stowell2015detection}. ASC is an emerging technology with potential applications in many fields. For example, smartphones could sense their environment in order to adapt their settings, e.g. using silent mode when situated in a concert hall\cite{barchiesi2015acoustic}. Hearing aids could adapt their filtering characteristics in changing scenarios, e.g. using beamforming when the user is in a conversation with one or multiple persons or giving an experience of surround sound during a music concert or when the user is on a street.

DCASE is a yearly international challenge where the detection or classification of acoustic scenes and events is an important re-occurring task\cite{mesaros2017detection, mesaros2018acoustic}. Within the last decade, plenty of methods have emerged in this field\cite{suh2020designing, hu2020device, sakashita2018acoustic, eghbal2016cp} and the classification accuracy on evaluation data that purely consists of recordings of acoustic scenes is continually improving. However, in many real life applications, there is no guarantee that the characterizing sound of an acoustic scene will be the dominant signal. Foreground speech might be captured as well, which is not always relevant towards the detection of the acoustic scene. For example, in telecommunications, or for audio captured by hearables, it is very likely that the captured audio is dominated by the speech of the user. In this case, existing ASC methods degrade significantly. Practical integration of ASC into the audio processing chain of such devices, therefore, requires robustness to foreground speech. 

Whereas research along this direction has not been actively pursued in the previous years, there exists a plethora of scientific work on the {\em denoising} of captured audio for the enhancement of foreground speech. For single-microphone captures (the focus in this paper), such denoising approaches typically take the form of a time-frequency gain function, which suppresses regions dominated by the background noise while preserving the speech-dominant regions. An overview of {\em classical} speech enhancement algorithms, based on statistical signal models, may be found in~\cite{hendriks13_nsoverview}. These have been surpassed in the recent past by model-based approaches and approaches based on deep-learning (e.g.,~\cite{mowlaee11_modelbasedenhancement,joder13_nmfNoise,mirsamadi16_dnnstat,yang18_dnnAR,elshamy2019dnn,strake2020} to cite but a few). Consequently, it seems that a logical approach to making ASC robust to foreground speech would be to apply a gain function which preserves the background and attenuates the {\em foreground speech}. Such a gain function maybe derived from the speech-enhancement gain. Other alternatives that suggest themselves are: the use of a Voice-Activity Detector (VAD) to extract features only in speech-absent frames or, as recently proposed in~\cite{liu2020towards}, an explicit first stage to {\em remove foreground speech}. 

However each alternative brings attendant disadvantages: constructing a gain function to remove speech while preserving the background leads to (non-linear) artefacts (e.g., due to estimation errors) in the resulting signal spectrum, which harms rather than improves the ASC backend. Using a VAD flag to estimate features only on speech-absent frames would lead to a long processing delay if foreground speech remains continually active. And, methods to explicitly remove foreground speech may be quite complex and may, again, introduce artefacts in the resulting signal. We propose, therefore, to investigate the feasibility of the noise-floor {\em estimate}, as an input stream for the ASC system. Such an estimate can be obtained by statistical methods with low complexity and in a real-time manner - thereby lending itself to practical applications.

Keeping complexity and low resource training in mind, we consider an ASC-backend based on a statistical framework. The winning contribution of the DCASE 2016~\cite{mesaros2016tut} challenge was the {\em binaural} iVector-based approach~\cite{dehak2010front}, described in~\cite{eghbal2016cp}. The system has a classification accuracy of 88.7\%, (an improvement of 11.5\% over the GMM-based baseline approach). In contrast, a deep convolutional neural network reported in~\cite{eghbal2016cp} achieves 83.3\% accuracy, indicating that the iVector approach is a robust choice in this relatively low-resource setting. Thus, our baseline shall be a competitive (in-house) iVector solution, on which we first investigate the effect of foreground speech.

Next, we extract the noise-floor from the input signal and derive the acoustic features for ASC from this. We consider statistical methods for estimating the noise-floor, and we refer the reader to~\cite{martin2001noise, rangachari2006noise,Gerkmann2012_mmse} for several well-known approaches. We finally contrast the robustness of the noise-floor based ASC system against our baseline. We also investigate the benefit of multi-condition training, in which foreground speech is added to the signals during the training process.

The remainder of this paper is organized as follows. \C{In Section~\ref{sec:prevWork}, we ouline the scope of this contribution.} In Section~\ref{sec:baseline} we describe our iVector baseline system where a modified Gaussian backend classifier with class-specific covariance models is adopted. In Section~\ref{sec:proposedMethod}, we (briefly) discuss the noise-floor estimation (based on~\cite{Gerkmann2012_mmse}) and its integration in the acoustic feature extraction of the iVector system. Section~\ref{sec:multiCondition} describes the multi-condition training. In Section~\ref{sec:exptSetup}, the dataset and experimental setup is described. In Section~\ref{sec:results}, we evaluate the performance in different settings with varying levels of foreground speech. Section~\ref{sec:conclusion} concludes the paper.

\C{
\section{Scope of paper}
\label{sec:prevWork}
The winning contribution of the DCASE 2016 Task 1 challenge~\cite{mesaros2016tut} was the {\em binaural} iVector-based approach~\cite{dehak2010front}, described in~\cite{eghbal2016cp}. The system exploits the information in both audio channels and has a classification accuracy of 88.7\%, (an improvement of 11.5\% over the baseline (GMM-based) approach). In contrast, a deep convolutional neural network reported in~\cite{eghbal2016cp} achieves 83.3\% accuracy, indicating that the iVector approach is a robust choice in this relatively low-resource setting. Our goal is to construct a competitive iVector solution and analyze its performance in the presence of foreground speech.  
Further, to enhance the robustness , we reduce the impact of the dominant foreground speech by extracting the targeted background signal through the construction of the noise-floor. Several well known approaches of noise-floor estimation can be found in~\cite{martin2001noise, rangachari2006noise, gerkmann2011noise,Gerkmann2012_mmse}.

Note that in this paper we will focus on single channel ASC. I.e., we do not perform the system analysis on the binaural data as it is difficult to add foreground speech to multi-channel audio in a realistic way without detailed information of the original recording setup. We will use the algorithm described in \cite{gerkmann2011noise} to extract the noise-floor in a statistical manner from single channel audio recordings.
}

\section{ASC iVector framework}
\label{sec:baseline}

We rely on the iVector framework to enable the processing of audio recordings by relatively simple backend classifiers. Through factor analysis the information in the variable length recording described by a sequence of MFCC feature vectors is compressed to a fixed length representation called the iVector. A Gaussian backend classifier, which fits a simple Gaussian model to the iVectors of each class, is then used to extract the most likely acoustic scene.
\C{The workflow diagram of the iVector baseline system is depicted in Fig.~\ref{diagram}.
\begin{figure}[ht]
\centerline{\includegraphics[width=7cm]{diagram.png}}
\caption{Workflow diagram of the iVector baseline system}
\label{diagram}
\end{figure}}

\subsection{Feature Extraction}
\label{subsec:feature extraction}
The MFCC features are constructed from the signal amplitude spectrum extracted with a frame length of 40 ms and 50\% overlap between successive frames. We extract 40 mel filter bank energies. 21 cepstral coefficients are kept, including the zeroth cepstral coefficient $c_0$. Normalization of this $c_0$ spectrum mean energy component did not have a significant performance impact during experiments on the DCASE 2016 Task 1 dataset.

Instead of adding first and second order derivatives to the MFCC features, we use Shifted Delta Cepstral (SDC) coefficients to obtain dynamic features with a larger temporal context\cite{torres2002approaches}. At frame index $n$, a total of $2K+1$ $\Delta$-feature vectors (first order time differences of the first $N$ static features) for the current frame and $K$ frames before and after the current frame are concatenated to the static MFCC coefficients. The considered $2K+1$ frames are separated by shifts of $P$ frames. Thus, a vector of length $(2K+1)\times N$ is added to each feature vector.

\subsection{Factor Analysis}
The variability in the observed MFCC features is caused by multiple underlying factors. To capture the most significant hidden properties of a recording such as the acoustic scene, we extract a fixed-length iVector~\cite{dehak2010front} for each recording.
The iVector framework uses an Universal Background Model (UBM) as a reference model and expresses for each recording how it deviates from this UBM. The UBM is a Gaussian Mixture Model (GMM) that is trained on a wide variety of data and deviations from this model are analyzed in the supervector domain. This domain is obtained by concatenating the GMM mean vectors of each component into a large supervector. The supervector $\bm{s}$ of a recording is approximated by adding a supervector that lays in the total variability subspace defined by a low-rank matrix $\bm{T}$ to the UBM supervector $\bm{m}$: 
\begin{equation}
    \bm{s}=\bm{m}+ \bm{T}\bm{w}.
    \label{gmm}
\end{equation}
$\bm{T}$ is called the iVector extractor, and the corresponding supervector subspace should cover most of the important variability observed in the data. The maximum a posteriori estimate $\bm{w}$ of the coordinates when enforcing a standard normal prior is called the iVector. $\bm{w}$ contains all relevant information of a recording, described by a fixed-length vector. Technical details of the iVector extraction process can be found in~\cite{glembek2011simplification}.
The unsupervised training process of the total-variability extractor $\bm{T}$ is initiated by Principal Component Analysis (PCA), followed by an Expectation-Maximization (EM) algorithm~\cite{glembek2011simplification}. 

\subsection{Gaussian Backend Classifier}
The estimated iVectors capture all kinds of variability, including within-class variability in the context of ASC. A simple Gaussian Backend (GB) suppresses this variability and focuses on inter-scene variability to perform the classification.

The standard GB approach fits a multivariate normal distribution to each class. For regularization purposes the common full covariance matrix $\bm{\Sigma}_s$ is shared between all classes and it is estimated by taking the unweighted average across all within-class covariance matrices $\bm{\Sigma}_c$. The class model that produces the maximum log-likelihood corresponds with the predicted acoustic scene. The simplified decision function $d(\bm{w})$ is
\begin{equation}
    d(\bm{w})=\underset{c}{\argmax} \left\{-\frac{1}{2} {(\bm{w}-\bm{\mu}_c)}^T \bm{\Sigma}_{s}^{-1}(\bm{w}-\bm{\mu}_c) \right\}
    \label{gb}
\end{equation}
with $\bm{\mu}_c$ the estimated mean iVector of acoustic scene class $c$. 

The hypothesis that the covariance matrix can be shared across all classes is reasonable in fields where the intra-class variability should be very similar, e.g.\ speaker variability in language or emotion recognition in speech. However, in the case of ASC the intra-class variability might be more class-specific and we propose to use class-dependent covariance matrices instead. The risk of overfitting is increased due to the larger number of model parameters and we decide to keep the regularization effect by sharing knowledge between classes during the covariance estimation. The estimated covariance matrix for class $c$ is defined by: $\widetilde{{\bm\Sigma}_c} = \alpha \bm{\Sigma}_{s} +(1-\alpha) {\bm\Sigma}_c$,
in which $\alpha \in [0,1]$ controls the regularization effect. In this case, the decision function $d(\bm{w})$ becomes
\begin{equation}
   \underset{c}{\argmax}\left\{ -\frac{1}{2} \log \vert{\widetilde{\bm{\Sigma}_c}}\vert -\frac{1}{2} {(\bm{w}-\bm{\mu}_c)}^T {\widetilde{\bm{\Sigma}_c}}^{-1}(\bm{w}-\bm{\mu}_c) \right\}
\end{equation}

\section{Noise-floor based features for ASC}
\label{sec:proposedMethod}

{
\C{In the case of ASC, non-stationary noise is present, such as babble noise and a car horn. Thus, voice activity detector (VAD) dealing with stationary noise is not available.}
The signal model for single-channel speech enhancement is typically formulated in the Short-Time Fourier Transform (STFT) domain as:
\begin{equation}
    X(\ell,k) = S(\ell,k) + N(\ell,k)\,,
\end{equation}
where $S(\ell,k)$ is the STFT representation of the foreground speech, $N(\ell,k)$ is the background and $X(\ell,k)$ is the captured (microphone) signal. In this representation $\ell$ is the discrete-frequency index and $k$ is the frame index. In the following discussion, the time- and frequency-indices will be omitted for brevity.

Statistical methods for noise-floor estimation are based on the following assumptions: (a) the foreground speech and background noise may be modelled as independent random variables and (b) the second-order statistics of the background noise is stationary over a longer period than the foreground speech. Essentially, the noise-floor estimate is obtained by recursively smoothing an estimate of the noise periodogram $\mathbb{E}\big(|N|^2\big\rvert X\big)$, thereby removing the short variations attributable to speech and preserving the long-term characteristics of the background. In order to efficiently track a varying noise-floor,~\cite{Gerkmann2012_mmse} proposed the use of an {\em adaptive} approach to the estimation of $\mathbb{E}\big(|N|^2\big\rvert X\big)$, driven by the conditional probability of speech presence given the observed signal.
Denoting the conditional probability of the hypothesis that speech is present by $P\big(\mathcal{H}_1\big\rvert X\big)$ and the probability of the alternative hypothesis by $P\big(\mathcal{H}_0\big\rvert X\big) = 1-P\big(\mathcal{H}_1\big\rvert X\big)$, we may obtain the following estimate of the noise periodogram:
\begin{equation}\label{eq:noiper}
    \noiper = (1-P\big(\mathcal{H}_1\big\rvert X\big))|X|^2 + P\big(\mathcal{H}_1\big\rvert X\big)\widehat{\noiflr}\,,
\end{equation}
where $\noiflre$ represents the noise-floor estimate at the previous time-frame ($\noiflre = \noiflre\tm$). The estimated noise periodogram from~\eqref{eq:noiper} is subsequently recursively smoothed with a fixed smoothing factor to obtain the updated noise-floor estimate $\noiflre$ for the current frame.

In the case of noise-floor based MFCC feature extraction, the acoustic features are now derived from the noise-floor estimate $\noiflre$. The rest of the feature extraction process remains the same compared to section \ref{subsec:feature extraction}. These MFCCs are used during training and evaluation of the proposed system.}

\section{Multi-condition training}
\label{sec:multiCondition}
For both the baseline and the proposed system using the noise-floor features, we also explore multi-condition training to further reduce the mismatch between training and testing conditions. This should result in better robustness of the classification system. A common way to measure the relation between the speech power and the noise power in a recording is via the Signal-to-Noise Ratio (SNR).
In this typical context, we refer to `noise' as the unwanted signal. In the case of ASC, the definition of `noise' is ambiguous, as the noise in the recording is useful for classification. Instead, we use the Speech-to-Background Ratio (SBR). To calculate the SBR, we need to estimate the speech and background power level. For speech, we use the active speech level of the speech fragment, for the background we use the Root Mean Square (RMS) value. In the proposed multi-condition training, mixed training data of different SBRs is used as training corpus compared to default of only using the original DCASE background noise.

\begin{table*}
\centering
\caption{Impact of the noise-floor features and multi-condition training on DCASE 2016 ASC accuracy for different SBRs.}
\label{table_accuracy}

\begin{tabular}{l|c|c|c|c|c|c|c|c}
\hline
\multicolumn{2}{c|}{\diagbox{Development dataset}{Accuracy(\%)}{Evaluation dataset} } & No speech & \dbu{-5} & \dbu{0} & \dbu{5} & \dbu{10} & \dbu{15} & \dbu{20} \\ \hline\multirow{2}{*}{\{No speech\}}            & Without noise-floor & \textbf{81.3}  & 62.3 & 59.2 & 55.1  & 50.5  & 44.9 &40.0  \\ \cline{2-9}
                                           & With noise-floor    & 76.2     & 74.4  & 70.8 & 67.4 & 63.3  & 57.4 & 52.8  \\ \hline
\multirow{2}{*}{\{No speech, \dbu{-5}\}}        & Without noise-floor & 79.2     & \textbf{77.2}  & \textbf{77.7} & \textbf{76.7} & 70.3 & 61.5  & 47.7  \\ \cline{2-9}
                                           & With noise-floor    & 76.2      & 75.1  & 74.6 & 71.0 & \textbf{71.0}  & \textbf{70.3}  & \textbf{66.2}   \\ \hline
\end{tabular}
\end{table*}
\begin{table*}
\centering
\caption{Impact of class-dependent Gaussian backend covariance models on DCASE 2016 ASC accuracy for different SBRs.}
\label{table_gb}
\begin{tabular}{l|c|c|c|c|c|c|c}
\hline
\diagbox{Development dataset}{Accuracy(\%)}{Evaluation dataset}  & No speech & \dbu{-5} & \dbu{0} & \dbu{5} & \dbu{10} & \dbu{15} & \dbu{20} \\ \hline
Standard Gaussian Backend (GB) & 74.9  & 70.3 & 68.7 & 66.4  & 62.1  & \textbf{57.7} &  48.2\\ \hline
Class-dependent GB covariance models  & \textbf{76.2}     & \textbf{74.4}  & \textbf{70.8} & \textbf{67.4} & \textbf{63.3}  & 57.4 & \textbf{52.8}   \\ \hline
\end{tabular}
\end{table*}

\section{Experimental setup} 
\label{sec:exptSetup}
\subsection{Dataset}
The proposed ASC system and foreground speech compensation techniques are evaluated on the DCASE 2016 Task 1 dataset\cite{mesaros2016tut}. This dataset contains 15 different acoustic scenes. The development dataset consists of almost 10\,hours of data. For each acoustic scene, 39\,minutes of data is provided, which is divided into 78 segments with a duration of 30\,seconds. The development data is split into four folds to enable cross-validation for hyperparameter tuning. The evaluation dataset contains approximately 3\,hours of data. Each acoustic scene has 13\,minutes of audio, divided into 26 segments of 30\,seconds. Recordings were made using a binaural microphone and the sampling rate is 44.1\,kHz. Our motivation in selecting this dataset is to validate our approaches for a situation where only limited {\em labelled} training data is available, which is commonly the case in bespoke industry scenarios. 

To simulate the presence of foreground speech we mix the recordings with speech from the pitch-tracking database from Graz University of Technology\cite{Pir_11} and the Multilingual Speech Database from NTT-AT\footnote{https://www.ntt-at.com/product/speech2002/}.
We downsampled the complete DCASE dataset to 16\,kHz mono to allow the use of these speech datasets, and to reduce the computational load during simulations. 
We make sure that there is no overlap between the speaker set used for mixing the training data and the speaker set used on the evaluation data.

\subsection{Setup}

All systems operate on MFCC features extracted by the procedure described in Section~\ref{subsec:feature extraction}. The temporal context is increased in the form of SDCs with parameters $M = 2$, $K = 2$, $N = 11$ and $P = 3$. The number of components in the UBM is 256 and the rank of the iVector extractor $T$ is set to 150. A weight $\alpha = 0.7$ is used to regularize the Gaussian backend with class-dependent covariance matrices. When the noise-floor estimation is applied, it is adopted for both the development and the evaluation dataset. All hyper-parameters are determined by four fold cross-validation on the development set.

The performance of all systems is expressed by the classification accuracy on the evaluation set, which is sufficient as a global performance measure as the DCASE 2016 dataset is balanced.

\section{Results and discussion} 
\label{sec:results}
In this section, we evaluate the impact of the noise-floor based MFCC features, the multi-condition training and the introduction of class-dependent covariance matrices in the Gaussian backend.

\subsection{Baseline System in Clean Conditions}
For a fair comparison we need to start from a competitive baseline system. The best reported iVector based result in the DCASE 2016 Task 1 challenge is 88.7\%~\cite{eghbal2016cp} on the 44.1\,kHz binaural evaluation data. Our proposed baseline system with class-dependent covariance matrices in the GB achieves 88.2\% on the same data. Similar to~\cite{eghbal2016cp}, the binaural data is processed by considering four possible input combinations of the audio channels (L, R, L+R and L-R) and applying score fusion by averaging the four different log-likelihood outputs. The achieved performance is close enough to the results reported in~\cite{eghbal2016cp} to do a fair study of the robustness of the system against active foreground speech. Adding foreground speech to the evaluation data is the most straightforward on mono 16\,kHz audio. If we consider a single audio channel, the performance of our baseline system drops to 85.4\%. Reducing the sample rate to 16\,kHz further reduces the accuracy to 81.3\%.

\subsection{Speech Robust ASC}
\begin{figure}[h!]
\centerline{\includegraphics[width=8.6cm]{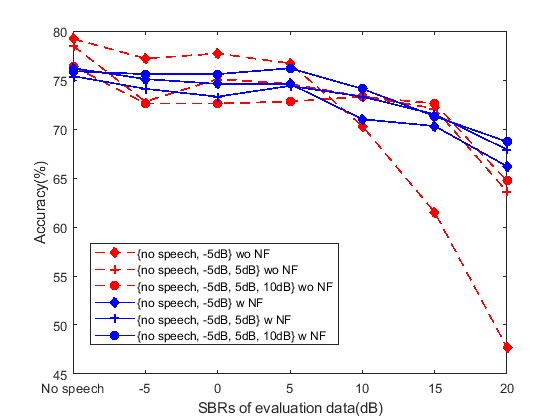}}
\vspace*{-2ex}
\caption{Impact of multi-condition training on the proposed systems. NF stands for the use of noise-floor based MFCC features.}
\label{multi}
\end{figure}

The performance of the proposed techniques to increase the robustness against foreground speech is shown in Table~\ref{table_accuracy}. The first two rows represent the results when the original DCASE development data is used for training and no speech is added. Different speech levels are added on top of the DCASE evaluation dataset with different SBRs of \{-5, 0, 5, 10, 15, 20\}\,dB. For each testing condition the SBR level is kept constant. In the first row, the reported accuracy of the iVector system on the standard MFCC features without noise-floor clearly deteriorates for increasing SBR levels when the foreground speech becomes increasingly prominent. The original accuracy of 81.3\% halves to 40\% in the \dbu{20} SBR condition. As shown in the second row, this degradation is {\em reduced significantly} when using the noise-floor based MFCC features. The performance is more consistent and, in the most adverse \dbu{20} SBR condition, the system now obtains 52.8\% accuracy.
\C{ A more detailed picture of the performance in this condition is given by the confusion matrix shown in Fig.~\ref{confusion}.} 
However, the performance in clean conditions slightly degraded to 76.2\%. This small degradation is to be expected as the noise-floor estimation will unavoidably remove some information that is useful for ASC. In certain real life scenarios, where the recordings can be strongly dominated by foreground speech rather than noise, the use of noise-floor based MFCC features should be preferred.
The final two rows in Table~\ref{table_accuracy} show the impact of Multi-Condition Training (MCT) for both standard and noise-floor based MFCC feature extraction. The training corpus now consists of the original DCASE development data and the development data augmented with speech with an SBR of \dbu{-5}. Similarly to the use of noise-floor based features, the use of multi-condition training reduces the negative impact of foreground speech. Using the default MFCC features with MCT  outperforms the standard training protocol with noise-floor based MFCCs, but its advantage tapers off for increasing SBR levels. For \dbu{20} SBR it is still outperformed by the noise-floor technique. We can increase the strength of the augmentation in MCT as shown in Fig.~\ref{multi}, which presents the results for \{no speech, \dbu{-5}\}, \{no speech, \dbu{-5}, \dbu{5}\}, and \{no speech, \dbu{5}, \dbu{-5}, \dbu{10}\} MCT training corpuses. But, it may be seen that  the increased performance for high SBR levels again comes at the cost of decreased performance on the original clean DCASE data. The final row of Table~\ref{table_accuracy} and Fig.~\ref{multi} show that improvements of noise-floor based MFCCs with MCT are complementary for higher SBR levels. The combination of both techniques (noise-floor based features and MCT) levels (flattens) the performance across a very wide range of SBR levels and is recommended if the encountered SBR levels in the real life applications are unpredictable.
\subsection{Gaussian Backend with Class-dependent Covariances}
The impact of introducing class-dependent covariance models in the GB is now evaluated on a wide range of SBR levels, including the clean evaluation data. The classification accuracies of the iVector system trained on noise-floor based MFCCs extracted from the original DCASE development data are shown in Table~\ref{table_gb}. The GB classifier with class-dependent covariance models consistently outperforms the standard GB on the whole range of SBR levels (except for SBR=\dbu{15} where both systems have similar performance). This indicates one has to be careful with the hypothesis that the intra-class variabilities are similar, especially in the context of ASC.

\vspace*{-3ex}
\vspace*{-2ex}

\section{Conclusions}
\label{sec:conclusion}
This paper presents two complementary techniques to increase the robustness of an ASC system against foreground speech: the incorporation of noise-floor based features and Multi-Condition Training (MCT). \C{The iVector based ASC system is enhanced by incorporating noise-floor based MFCCs and Multi Condition Training.} The noise-floor based features can successfully extract the acoustic scene in the presence of foreground speech, but MCT still helps in eliminating any mismatch between training and testing conditions. The combination of both techniques achieves very consistent DCASE 2016 classification performance that is {\em almost independent} of the energy level of the encountered nuisance speech. An absolute improvement of up to 25\% is seen in the most adverse conditions. Future work will verify if the noise-floor can be used for feature extraction and/or for data augmentation in neural network based ASC solutions.




\bibliographystyle{IEEEtran}
\bibliography{ref.bib}
\end{document}